\documentclass[aps,preprint,tightenlines,showpacs]{revtex4-1}
\usepackage{amssymb}

\usepackage{dcolumn}
\usepackage{bm}

\begin{document}

\title{Gauge Theory of Maxwell-Weyl Group}
\author{O. Cebecio\u{g}lu$^1$ \footnote{%
E-mail: ocebecioglu@kocaeli.edu.tr} and S. Kibaro\u{g}lu$^1$ \footnote{%
E-mail: salihkibaroglu@gmail.com}}

\date{\today}

\begin{abstract}
Starting from Maxwell-Weyl algebra we found the transformation rules for
generalized space-time coordinates and the differential realization of
corresponding generators. By treating local gauge invariance of Maxwell-Weyl
group, we presented the Einstein-Cartan-Weyl gravity with the additional
terms containing the gauge fields associated with the antisymmetric
generators.
\end{abstract}

\affiliation{$^1$Department of Physics, Kocaeli University, 41380 Kocaeli,
Turkey}

\pacs{02.20.Sv; 04.20.Fy; 11.15.-q; 11.10.Ef}

\maketitle

\section{Introduction}

Symmetries are of great importance in the formulation of theories describing
a given physical system. Invariance of any physical system under a given
symmetry transformation group determines its properties to a great extent.
In order to get more information on the physical systems one has to find new
symmetries or enlarge the symmetries of the system. For instance, there is a
well known theorem, the Coleman-Mandula No-Go theorem \cite{CM} says that
there is no non-trivial way to combine space-time and internal symmetries.
In this case symmetry generators are bosonic and their Lie algebra includes
only commutators. But there is one non-trivial way to get around this by
incorporating anticommutators for fermionic generators. This type of
extension leads to supersymmetry. There is a different extension of Poincare
algebra bypassing another well known theorem \cite{G} that does not allow
non-central extension of this algebra. In this case one introduces a new
anti-symmetric tensor generator by imposing non-commutativity of momentum
generators and satisfying $\left[P_{a},P_{b}\right] =iZ_{ab}$. Motivation
behind this kind of extension is that symmetries of empty Minkowski
space-time is described by Poincare algebra if such a space-time filled with
some background field must lead to modification of the Poincare algebra. The
extension of the Poincare algebra by six additional abelian generators is
called the Maxwell algebra \cite{BCR1,BCR2,SCH,BH,NO1,NO2}. Addition of new
generators to the Poincare algebra leads naturally to extended space-time
geometry.

A decade ago, another possible tensor extension of Poincare algebra was
proposed and its supersymmetric generalization was studied in \cite{SS1},
and two years later semi-simple extension and also supersymmetric
generalization of Poincare algebra were presented in \cite{SS2}. After these
articles, Maxwell symmetries has appeared in literature once more. Different
deformation of Maxwell algebras, their supersymmetric generalization and
their dynamical realization for massless superparticle model were
investigated in \cite{BGKL1,GGP,BGKL2,BGKL3}. These extended symmetries were
also applied to the planar dynamics of the Landau problem and to the
description of higher spin fields \cite{FL2,FL3}. N-extended Maxwell algebra
was constructed via a contraction procedure of (extended Lorentz $\bigoplus $
extended AdS) superalgebras to arrive various type of N-extended Maxwell
superalgebras \cite{L,KK,KL,AILW}. Based on these kind of Maxwell
superalgebras, several Maxwell supergravity models were studied by the
following authors \cite{SS3,SS4,AI,DG2,HA}.

J.A. de Azcarraga in his paper \cite{AZKL1} achived the generalized
cosmological constant problem by gauging the Maxwell algebra, later D.V.
Soroka \cite{SS4} presented another approach to the problem but this time
gauging semi-simple extension of Poincare algebra. The formulation of
different types of Maxwell gravities has already been done in \cite
{AZKL1,DG1,DG3}. With this motivation, we have presented the gauge theory of
Maxwell-Weyl group denoted by $\mathcal{M}\mathcal{W}(1,3)$.

The paper is organized as follows: In Sec. II we consider the Weyl algebra
and its extension by an antisymmetric tensor generator. Applying the method
of non-linear coset realization \cite{CWZ1,CCWZ1,SAS1,SAS2} to Maxwell-Weyl
group, the transformation rules for generalized coordinates are found and
explicit expression for the generators of the Maxwell-Weyl algebra are
given. In Sec. III we discussed local gauge theory based on the Maxwell-Weyl
algebra by introducing vierbein, spin connection, dilatation, six additional
geometric abelian gauge fields corresponding to the antisymmetric generator
and a compansating field in order to preserve Weyl scaling. From these we
deduced the transformations of the fields under local gauge transformation
and their covariant curvatures that leave the Lagrangian invariant. After
that we constructed invariant Lagrangian and finally, we obtained the
equations of motion for all dynamical variables.

\section{Weyl Algebra and Its Tensor Extension}

Scale transformation together with Poincare transformations form the eleven
dimensional Weyl group, $\mathcal{W}(1,3)$. Its Lie algebra is generated by
the operators $P_{a}$, $M_{ab}$ and $\ D$ associated with space-time
translations, restricted Lorentz transformations and dilatations,
respectively. The non-zero commutation relations can be written in the
following form: 
\begin{eqnarray}
\left[ M_{ab},M_{cd}\right] &=&i\left( \eta _{ad}M_{bc}+\eta
_{bc}M_{ad}-\eta _{ac}M_{bd}-\eta _{bd}M_{ac}\right)  \nonumber \\
\left[ M_{ab},P_{c}\right] &=&i(\eta _{bc}P_{a}-\eta _{ac}P_{b})  \nonumber
\\
\left[ P_{a},D\right] &=&iP_{a}
\end{eqnarray}
and differential realization of the generators are given by 
\begin{eqnarray}
M_{ab}&=&i(x_{a}\partial _{b}-x_{b}\partial _{a})  \nonumber \\
P_{a}&=&i\partial _{a}  \nonumber \\
D&=&ix^{a}\partial _{a}
\end{eqnarray}
where $\eta _{ab}$ is the Minkowski metric $\eta _{ab}=diag(+1,-1,-1,-1)$
with the tangent space indices $\ a, b$ run over $\ 0,1,2,3 $. The Weyl
algebra can be extended with antisymmetric tensor generator $Z_{ab}$ by the
use of expansion method introduced in \cite{AILW,AIPV,LC}, see also \cite
{GGP1,BG1,BG2}. One could also consider Maxwell-Weyl algebra as the
extension of Maxwell algebra by dilatation generator \cite{BGKL2}. Here we
present a novel 17-dimensional Maxwell-Weyl algebra with the following
commutation rules: 
\begin{eqnarray}
\left[ M_{ab},M_{cd}\right] &=&i\left( \eta _{ad}M_{bc}+\eta
_{bc}M_{ad}-\eta _{ac}M_{bd}-\eta _{bd}M_{ac}\right)  \nonumber \\
\left[ M_{ab},P_{c}\right] &=&i(\eta _{bc}P_{a}-\eta _{ac}P_{b})  \nonumber
\\
\left[ P_{a},P_{b}\right] &=&iZ_{ab}  \nonumber \\
\left[ P_{a},D\right] &=&iP_{a}  \nonumber \\
\left[ M_{ab},Z_{cd}\right] &=&i\left( \eta _{ad}Z_{bc}+\eta
_{bc}Z_{ad}-\eta _{ac}Z_{bd}-\eta _{bd}Z_{ac}\right)  \nonumber \\
\left[ Z_{ab},D\right] &=&2iZ_{ab}
\end{eqnarray}
and all the other commutators vanish identically.

We introduce the coordinates $X^{M}=(x^{a},\theta ^{ab}=-\theta ^{ba})$ as
the group space coordinates which are dual to $P_{a}, Z_{ab}$. One can
regard generalized space-time as a coset space of the Maxwell group with
respect to the Lorentz group \cite{FL1,BGKL2}. In the case of Maxwell-Weyl
symmetry, the variables $x^{a},\theta^{ab},\sigma $ parametrize the coset
element as 
\begin{equation}
K(x,\theta ,\sigma )=e^{ixp}e^{i\theta Z}e^{i\sigma D}
\end{equation}
The infinitesimal transformation of the coset parameters, generated by the
constant group elements $a,\varepsilon ,\lambda ,u$ corresponds to a motion
in tensorially extended space-time induced by the left multiplications 
\begin{equation}
g(a,\varepsilon ,\lambda ,u)K(x,\theta ,\sigma )=K(x^{\prime },\theta
^{^{\prime }},\sigma ^{^{\prime }})h(\omega )
\end{equation}
where 
\begin{equation}
h(\omega )=e^{-\frac{i}{2}\omega M}
\end{equation}
and can be easily evaluated through the use of the well known
Baker-Hausdorff-Campell formula:$\ $%
\begin{equation}
e^{A}e^{B}=e^{A+B+\frac{1}{2}\left[ A,B\right] }
\end{equation}
which holds when $\left[ A,B\right] $ commutes with both $A\ $and$\ B.$
Infinitesimally, the action of the $\mathcal{M}\mathcal{W}(1,3)$ on group
space coordinates reads 
\begin{eqnarray}
\delta x^{a}&=&{u^{a}}_{b}x^{b}+\lambda x^{a}+a^{a}  \nonumber \\
\delta \theta ^{ab}&=&\varepsilon ^{ab}-\frac{1}{4}a^{[a}x^{b]}+2\lambda
\theta ^{ab}+{u^{[a}}_{c}\theta ^{cb]}  \nonumber \\
\delta \sigma &=&\lambda  \nonumber \\
\omega ^{ab}&=&u^{ab}
\end{eqnarray}
where anti-symmetrization is defined by $\
A_{[a}B_{b]}=A_{a}B_{b}-A_{b}B_{a} $.

As seen from transformation rules we simply have Weyl transformation for
space-time coordinates and there is no contribution to that resulting from
tensorial extension. On the other hand as seen from transformation rules two
translation yields a shift of $\theta^{ab},\ $induces some additional terms
on tensorial space. The transformation law for a scalar field $\Phi ^{\prime
}\left( x^{a},\theta^{ab}\right) =\Phi \left( x^{a}-\delta x^{a},\theta
^{ab}-\delta \theta^{ab}\right) $ under the infinitesimal action of $%
\mathcal{M}\mathcal{W}(1,3)$ implies 
\begin{equation}
\delta \Phi =-\delta x^{a}\partial _{a}\Phi -\delta \theta ^{ab}\partial
_{ab}\Phi =i\left( a^{a}P_{a}+\varepsilon ^{ab}Z_{ab}+\lambda D-\frac{1}{2}%
u^{ab}M_{ab}\right) \Phi
\end{equation}
with 
\begin{eqnarray}
P_{a}&=&i(\partial _{a}-\frac{1}{2}x^{b}\partial _{ab})  \nonumber \\
Z_{ab}&=&i\partial _{ab}  \nonumber \\
D&=&i(x^{a}\partial _{a}+2\theta ^{ab}\partial _{ab})  \nonumber \\
M_{ab}&=&i\{x_{a}\partial _{b}-x_{b}\partial _{a}+2({\theta _{a}}%
^{c}\partial _{bc}-{\theta _{b}}^{c}\partial _{ac})\}
\end{eqnarray}
where $\partial _{a}=\frac{\partial }{\partial x^{a}}$, $\partial _{ab}=%
\frac{\partial }{\partial \theta ^{ab}}$ and one can check that these
generators fulfil the Maxwell-Weyl algebra, $\mathit{mw}(1,3)$, and satisfy
the all Jacobi identities.

\section{Gauging the Maxwell-Weyl Algebra}

Gauge theories of ordinary Weyl group were treated in \cite
{OM,BE,CT,KAS,BF,Blago}. In this section we consider gauge theory of the $%
\mathit{mw}(1,3)$ algebra. In order to gauge this algebra one introduces a
connection $
\mathcal{A}_{\mu }(x)$ with 
\begin{equation}
\mathcal{A}_{\mu }=e_{\mu }^{a}P_{a}+B_{\mu }^{ab}Z_{ab}+\chi _{\mu }D-\frac{%
1}{2}\omega _{\mu }^{ab}M_{ab}
\end{equation}
where $e_{\mu }^{a}(x)$ is the vierbein, $\omega _{\mu}(x)$ is the spin
connection, $B_{\mu }^{ab}(x)$ is the gauge field corresponding to the
antisymmetric tensor generator and $\chi _{\mu }(x)$ is the gauge field
corresponding to dilatation generator. The variation of these fields under
infinitesimal gauge transformations is given by 
\begin{equation}
\delta \mathcal{A}_{\mu }=-\partial _{\mu }\zeta -i\left[ \mathcal{A}_{\mu
},\zeta \right]  \label{delta_amu}
\end{equation}
with the gauge generator 
\begin{equation}
\zeta \left( x\right) =y^{a}\left( x\right) P_{a}+\varphi
^{ab}(x)Z_{ab}+\rho (x)D-\frac{1}{2}\tau ^{ab}(x)M_{ab}
\end{equation}
where $y^{a}\left( x\right) $ are space-time translations, $\varphi ^{ab}(x)$
are translations in tensorial space, $\rho (x)$ dilatation parameter and $%
\tau ^{ab}(x)$ are the Lorentz transformation parameters. From Maxwell-Weyl
algebra and Eq.(\ref{delta_amu}) it follows 
\begin{eqnarray}
\delta e_{\mu }^{a} &=&-\partial _{\mu }y^{a}-\omega _{\mu b}^{a}y^{b}-\chi
_{\mu }y^{a}+\rho e_{\mu }^{a}+\tau _{\ b}^{a}\ e_{\mu }^{b}  \nonumber \\
\delta B_{\mu }^{ab} &=&-\partial _{\mu }\varphi ^{ab}-\omega _{\mu c}^{[a}{%
\varphi ^{cb]}}-2\chi _{\mu }\varphi ^{ab}+\frac{1}{2}e_{\mu
}^{[a}y^{b]}+2\rho B_{\mu }^{ab}+{\tau ^{[a}}_{c}{\ B}_{\mu }^{cb]} 
\nonumber \\
\delta \chi _{\mu } &=&-\partial _{\mu }\rho  \nonumber \\
\delta \omega _{\mu }^{ab} &=&-\partial _{\mu }\tau ^{ab}-\omega _{\mu
c}^{[a}{\tau ^{cb]}}  \label{delta_e}
\end{eqnarray}
The curvature two-form $\mathcal{\digamma }$ is given by the structure
equation 
\begin{equation}
\ \mathcal{\digamma }\ =d\mathcal{A}+i\mathcal{A}\wedge \mathcal{A}=d%
\mathcal{A}+\frac{i}{2}\left[ \mathcal{A},\mathcal{A}\right] =\frac{1}{2}%
e^{a}\wedge e^{b}\mathcal{\digamma }_{ab}
\end{equation}
whence writing 
\begin{equation}
\mathcal{\digamma }=F^{a}P_{a}+F^{ab}Z_{ab}+fD-\frac{1}{2}R^{ab}M_{ab}
\label{f1}
\end{equation}
we find 
\begin{eqnarray}
F^{a} &=&de^{a}+{\omega ^{a}}_{b}\wedge e^{b}+\chi \wedge e^{a}  \nonumber \\
R^{ab} &=&d\omega ^{ab}+{\omega ^{a}}_{c}\wedge \omega ^{cb}  \nonumber \\
F^{ab} &=&dB^{ab}+{\omega ^{\lbrack a}}_{c}\wedge B^{cb]}+2\chi \wedge
B^{ab}-\frac{1}{2}e^{a}\wedge e^{b}  \nonumber \\
f &=&d\chi
\end{eqnarray}
The explicit expressions of curvatures defined by Eq.(\ref{f1}) are 
\begin{eqnarray}
F_{\mu \nu }^{a} &=&\partial _{\lbrack \mu }e_{\nu ]}^{a}+\omega _{\lbrack
\mu b}^{a}e_{\nu ]}^{b}+\chi _{\lbrack \mu }e_{\nu ]}^{a}  \nonumber \\
R_{\mu \nu }^{ab} &=&\partial _{\lbrack \mu }\omega _{\nu ]}^{ab}+\omega
_{\lbrack \mu c}^{a}\omega _{\nu ]}^{cb}  \nonumber \\
F_{\mu \nu }^{ab} &=&\partial _{\lbrack \mu }B_{\nu ]}^{ab}+\omega _{\lbrack
\mu c}^{[a}B_{\nu ]}^{cb]}+2\chi _{\lbrack \mu }B_{\nu ]}^{ab}-\frac{1}{2}%
e_{[\mu }^{a}e_{\nu ]}^{b}  \nonumber \\
f_{\mu \nu } &=&\partial _{\lbrack \mu }\chi _{\nu ]}
\end{eqnarray}
These are the torsion tensor, curvature tensor, a component corresponding to
the tensor generator $Z_{ab}$ and a component corresponding to the
dilatation generator $D$ respectively. 

Under an infinitesimal gauge transformation with parameters $\zeta $, the curvature 2-form $\ \mathcal{%
\digamma }$ transform as 
\begin{equation}
\delta \mathcal{\digamma }_{\mu \nu }=i\left[ \zeta ,\mathcal{\digamma }%
_{\mu \nu }\right]
\end{equation}
and hence one gets 
\begin{eqnarray}
\delta F_{\mu \nu }^{a} &=&-y^{a}f_{\mu \nu }-R_{\mu \nu b}^{a}y^{b}+\rho
F_{\mu \nu }^{a}+{\tau ^{a}}_{b}F_{\mu \nu }^{b}  \nonumber \\
\delta F_{\mu \nu }^{ab} &=&-2\varphi ^{ab}f_{\mu \nu }+{\varphi ^{[a}}_{c}R_{\mu \nu }^{cb]}-\frac{1}{2}y^{[a}F_{\mu \nu }^{b]}+2\rho F_{\mu
\nu }^{ab}+{\tau ^{[a}}_{c}F_{\mu \nu }^{cb]}  \nonumber \\
\delta f_{\mu \nu } &=&0  \nonumber \\
\delta R_{\mu \nu }^{ab} &=&{\tau ^{[a}}_{c}R_{\mu \nu }^{cb]}
\label{delta_F}
\end{eqnarray}

The field strengths Eq.(\ref{f1}) may be used to construct $\mathcal{MW}%
(1,3) $ invariant free Lagrangians for the corresponding gauge fields. To
construct Lagrangians which are locally invariant under the whole group,
special care must be given to the scale invariance and its relation to
dilatation subgroup of $\mathcal{MW}(1,3)$. Localization of the dilatation
symmetry bring us back to Weyl-gauge theory. One observes that infinitesimal
action of dilatation on vierbein induces a change on the metric tensor as 
\begin{equation}
\delta g_{\mu \nu }\left( x\right) =2\rho \left( x\right) g_{\mu \nu }\left(
x\right)  \label{g_munu}
\end{equation}
which is exactly the Weyl-gauge transformation and it considerably restricts
the form of any action built from curvature tensors. Eq.(\ref{g_munu}) means
that the metric tensor $g_{\mu \nu }\left( x\right) $ has scale(Weyl) weight
two i.e. $w\left( g_{\mu \nu }\right) =+2$, consequently the reciprocal $%
g^{\mu \nu }\left( x\right) $ has weight $-2$ and $\sqrt{-g}$ is of \ weight 
$+4$. From transformation rules Eq.(\ref{delta_e}) and Eq.(\ref{delta_F}),
we immediately infer that gauge fields $e_{\mu }^{a}$, $B_{\mu }^{ab}$, $%
\chi _{\mu }$, $\omega _{\mu }^{ab}$, and field strengths $F_{\mu \nu }^{a}$%
, $R_{\mu \nu }^{ab}$, $F_{\mu \nu }^{ab}$, $f_{\mu \nu }$have the following
Weyl weights $1$, $2$, $0$, $0$ and $1$, $0$, $2$, $0$ respectively.

After these preliminaries for Weyl weight we can obtain the Bianchi
identities corresponding to the curvature forms as 
\begin{equation}
\mathcal{D}R^{ab}=0
\end{equation}
\begin{equation}
\mathcal{D}F^{ab}=R{^{[a}}_{c}\wedge B^{cb]}+2f\wedge B^{ab}-\frac{1}{2}F{%
^{[a}}\wedge e^{b]}
\end{equation}
\begin{equation}
\mathcal{D}F^{a}=R{^{a}}_{b}\wedge e^{b}+f\wedge e^{a}
\end{equation}
where $\mathcal{D}\Phi =[d+\omega +w(\Phi )\chi ]$ $\Phi $ is the
Lorentz-Weyl covariant derivative with $\omega =-\frac{i}{4}\omega ^{\alpha
\beta }\Sigma _{\alpha \beta }$ and $w$ being Weyl weight of the corresponding
field.

The free gravitational action has the form 
\begin{equation}
S_{f}=\int d^{4}xe\mathcal{L}_{f}
\end{equation}
requiring that the action has Weyl weight zero. $w\left( \sqrt{-g}\right)
=w\left( e\right) =+4$, implies that the Lagrangian density must satisfy the
condition $w\left( \mathcal{L}_{f}\right) =-4.$ Since the scalar curvature
has $w\left( R\right) =-2$, it is not allowed to appear linearly in the
action. In order to have consistent theory of gravity Weyl used quadratic
terms in the action yielding the fourth order field equations \cite{Weyl}.
Clearly, $R$ by itself inappropriate, however if we multiply it by a
compensating scalar field introduced by Brans-Dicke \cite{BD} and elaborated
by Dirac \cite{Dirac,dereli,agnese} one can form Weyl invariant action linear in R.

In our approach we will follow Dirac's idea. The scalar field $\phi $ with
Weyl weight $-1$ let $\phi ^{2}R$ be regular part of $\mathcal{L}_{f}$ \ and
hence the following combination (shifted curvature) 
\begin{equation}
\mathcal{J}^{ab}=R^{ab}+2\gamma \phi ^{2}F^{ab}
\end{equation}
have Weyl weight zero. With this combination we have Einstein Lagrangian
that involves the curvature scalar linearly. Therefore we consider the
following Lagrangian density 4-form as our starting point for the free
gravitational part:

\begin{eqnarray}
L_{f} &=&\frac{1}{2\kappa \gamma }\mathcal{J}\wedge ^{\ast }\mathcal{J}=%
\frac{1}{4\kappa \gamma }\varepsilon _{abcd}\mathcal{J}^{ab}\wedge \mathcal{J%
}^{cd}  \nonumber \\
&=&\frac{1}{4\kappa \gamma }\varepsilon _{abcd}R^{ab}\wedge R^{cd}+\phi ^{2}%
\frac{1}{\kappa }\varepsilon _{abcd}R^{ab}\wedge F^{cd}+\phi ^{4}\frac{%
\gamma }{\kappa }\varepsilon _{abcd}F^{ab}\wedge F^{cd}
\end{eqnarray}
where $\gamma $ and $ \kappa $ are constants and the first term can be ignored because it is a closed form. Introduction of compensating field forces us to add its kinetic term to the Lagrangian by means
of full covariant derivative which in turn includes the gauge field $\chi_{\mu }$ hence one has to add one more Maxwell like kinetic term to Lagrangian. For completeness we can add to Lagrangian the further Weyl invariant self-interacting term $\frac{\lambda }{4}\phi ^{4}$ with $\lambda $ another constant. We then get the total action for vacuum as follows:

\begin{equation}
L_{o}=\frac{1}{4}f\wedge ^{\ast }f-\frac{1}{2}\mathcal{D}\phi \wedge ^{\ast }%
\mathcal{D}\phi +\frac{\lambda }{4}\phi ^{4}{}^{\ast }1
\end{equation}
Here  `*' the asterisk \ denotes the Hodge duality operation then our
complete action is the sum of the free gravity action
and the vacuum action:

\begin{eqnarray}
S &=&\int \phi ^{2}\frac{1}{\kappa }\varepsilon _{abcd}R^{ab}\wedge
F^{cd}+\phi ^{4}\frac{\gamma }{\kappa }\varepsilon _{abcd}F^{ab}\wedge F^{cd}
\nonumber \\
&&+\frac{1}{4}f\wedge ^{\ast }f-\frac{1}{2}\mathcal{D}\phi \wedge ^{\ast }%
\mathcal{D}\phi +\frac{\lambda }{4}\phi ^{4}{}^{\ast }1  \label{action}
\end{eqnarray}

Since the local translation together with tensorial translation being traded for diffeomorphism invariance are not symmetries of the action \cite{DG1}, omitting both local space-time and tensorial-space translations the transformation rules for the curvature can be rewritten as
\begin{eqnarray}
\delta F_{\mu \nu }^{a} &=&\rho F_{\mu \nu }^{a}+{\tau ^{a}}_{b}F_{\mu \nu
}^{b}  \nonumber \\
\delta F_{\mu \nu }^{ab} &=&2\rho F_{\mu \nu }^{ab}+{\tau ^{[a}}_{c}F_{\mu
\nu }^{cb]}  \nonumber \\
\delta f_{\mu \nu } &=&0  \nonumber \\
\delta R_{\mu \nu }^{ab} &=&{\tau ^{[a}}_{c}R_{\mu \nu }^{cb]}
\end{eqnarray}
Clearly the components of the shifted curvature transform in a homogeneous
way as 
\begin{equation}
\delta \mathcal{J}^{ab}={\tau^{[a}}_{c}\mathcal{J}_{\mu \nu }^{cb]}
\end{equation}
since the transformation rule for the scalar field is $\delta \phi =-\rho \phi.$ This ensures that the free gravity part of the action is gauge invariant besides that the vacuum part is already gauge invariant by construction.

Invariance under local Maxwell-Weyl transformations (diffeomorphism) can be directly checked
by using the explicit form of the Lie derivative. Since 
\begin{equation}
\ \delta S_{diff}=\int l_{\xi }L=\int \left( di_{\xi }L+i_{\xi }dL\right)
\end{equation}
the first term is a total divergence which can be ignored as a surface
integral and the second term is zero since the $5-$form $dL$ vanishes
identically on the \ $4-$dimensional space-time hence $\delta S=0$.

In order to show the diffeomorphism invariance of the action explicitly, one substitudes the transformation rules Eq.(\ref{delta_e}) to the following action integral
\begin{equation}
\ \delta S=\int \delta e^{a}P_{a} + \delta B^{ab}V_{ab} +\delta \chi Q+ \delta\phi U +\delta \omega^{ab}Y_{ab}
\end{equation}
then one gets the following identities (conservation rules)

\begin{eqnarray}
\mathcal{D}P_{a}-V_{ab}e^{b}&=&0 \nonumber \\
\mathcal{D}V_{ab}&=&0 \nonumber \\
e^{a}P_{a}+2B^{ab}V_{ab}+\mathcal{D}Q&=&0 \nonumber \\
U&=&0 \nonumber \\
e_{a}P_{b}-B_{[ad}{V^{d}}_{b]}-\mathcal{D}Y_{ab}&=&0
\end{eqnarray}

The field equations are  $ P_{a}=0 $, $ V_{ab}=0 $, $ Q=0 $, $ U=0 $ and $ Y_{ab}=0 $ or they can found by varying the action Eq.(\ref{action}) directly
over the independent variables $\omega $, $e$, $B$, and $\varkappa $.
Varying over the connection components $\omega $ we obtain the following
equation for the generalized torsion tensor 
\begin{equation}
\mathcal{D}\left( \phi ^{2}F^{ab}\right) -\phi ^{2}{\mathcal{J}^{[a}}
_{c}\wedge B^{cb]}=0
\end{equation}
The variation of the action Eq.(\ref{action}) with respect to $e$ leads to 
\begin{eqnarray}
&&-\phi ^{2}\frac{1\text{\ }}{\kappa }\varepsilon _{abcd}\mathcal{J}%
^{ab}\wedge e^{d}+\frac{1}{2}[\mathcal{D}_{c}\phi ^{\ast }\mathcal{D}\phi +%
\mathcal{D}\phi \wedge ^{\ast }\left( e_{a}\wedge e_{c}\right) \mathcal{D}%
^{a}\phi ]  \nonumber \\
&&-\frac{1}{4}(f_{cb}e^{b}\wedge ^{\ast }f-\frac{1}{2}\varepsilon
_{abcd}f^{ab}e^{d}\wedge f)+\frac{\lambda }{4}\phi ^{4}\text{ }^{\ast
}e_{c}=0 \label{e_var}
\end{eqnarray}
Furthermore, the $B$ variation of Eq.(\ref{action}) gives 
\begin{equation}
\mathcal{D}(\phi ^{2}\mathcal{J}^{ab})=0
\end{equation}
while the $\chi $ variation of Eq.(\ref{action}) yields the following
equations of motion 
\begin{equation}
\frac{2\phi ^{2}}{\kappa }\epsilon _{abcd}\mathcal{J}^{ab}\wedge B^{cd}+%
\frac{1}{2}\mathcal{D}^{\ast }f+\phi ^{\ast }\mathcal{D}\phi =0.
\end{equation}
and finally we obtain equation of motion for the scalar field $\phi $: 
\begin{equation}
\frac{2\phi }{\kappa }\epsilon _{abcd}\mathcal{J}^{ab}\wedge F^{cd}+\mathcal{%
D}^{\ast }\mathcal{D}\phi +\lambda \phi ^{3}{}^{\ast }1=0.
\end{equation}

The equations obtained are invariant with respect to the local Maxwell-Weyl
transformation considered above. By a straightforward calculation, one can
show that all these equations of motion verify each other. In terms of
shifted curvature, Eq.(\ref{e_var}) becomes
\begin{eqnarray}
\phi ^{2}\left( {\mathcal{J}^{a}}_{b}-\frac{1}{2}{\delta ^{a}}_{b}\mathcal{J}%
\right) &=&-\frac{\kappa }{2}[\mathcal{D}^{a}\phi \mathcal{D}_{b}\phi -\frac{%
1}{2}{\delta ^{a}}_{b}\left( \mathcal{D}_{c}\phi \mathcal{D}^{c}\phi -\frac{%
\lambda }{2}\phi ^{4}\right) + \\
&&+\left( f^{ac}f_{cb}-\frac{1}{4}{\delta ^{a}}_{b}f_{cd}f^{cd}\right) ]
\end{eqnarray}
and passing from tangent space to world indices one gets field
equation with cosmological term depending on dilaton field 
\begin{equation}
{R^{\mu }}_{\alpha }-\frac{1}{2}{\delta ^{\mu }}_{\alpha }R-3\gamma \phi ^{2}%
{\delta ^{\mu }}_{\alpha }=2\gamma \phi ^{2}{T}\left( B\right) {^{\mu }}%
_{\alpha }-\frac{\kappa }{2}\phi ^{-2}\left[ {T}\left( \phi \right) {^{\mu }}%
_{\alpha }+{T}\left( f\right) {^{\mu }}_{\alpha }\right]
\end{equation}
where 
\begin{equation}
{T}\left( B\right) {^{\mu }}_{\alpha }=e_{a}^{\mu }e_{b}^{\beta }{}\mathcal{D%
}_{[\alpha }B_{\beta ]}^{ab}-\frac{1}{2}{\delta ^{\mu }}_{\alpha }\left(
e_{a}^{\rho }e_{b}^{\sigma }{}\mathcal{D}_{[\rho }B_{\sigma ]}^{ab}\right)
\end{equation}
\begin{equation}
{T}\left( \phi \right) {^{\mu }}_{\alpha }=\mathcal{D}^{\mu }\phi \mathcal{D}%
_{\alpha }\phi -\frac{1}{2}{\delta ^{\mu}}_{\alpha}\left( \mathcal{D}%
_{\gamma }\phi \mathcal{D}^{\gamma }\phi -\frac{\lambda }{2}\phi ^{4}\right)
\end{equation}
\begin{equation}
{T}\left( f\right) {^{\mu }}_{\alpha }=f^{\mu \beta }f_{\beta \alpha }-\frac{%
1}{4}{\delta ^{\mu}}_{\alpha}f_{\gamma \delta }f^{\gamma \delta }
\end{equation}
are the energy-momentum tensors for the $B$-gauge field, dilaton field and $%
\chi $-field respectively. 

It is an interesting exercise to rewrite the above system of equations in
terms of torsion-free connection. In such a reformulation, the torsional effects due to the scalar field coupling to gravity may be interpreted as an additional contribution to the stress energy forms.

\section{Conclusion}

In the present paper we considered the algebra of generators and constructed
a non-linear realization of the Maxwell-Weyl group on its coset space with
respect to the Lorentz group. We proposed here the gauge theory formulation
of Maxwell-Weyl gravity in order that we introduced the corresponding gauge
fields, presented field transformations and found the equations of motion.
Defining scale invariant shifted curvature, we achieved extension of Einstein-Cartan-Weyl field equation with
variable cosmological term and additional source term. From source term we derived
the stress energy-momentum tensor and from there we concluded that
introduction of dilatation to the theory does affect the cosmological
constant and contributes to energy-momentum of the B-field.

Appearance of the cosmological constant term as a dynamical variable in the
presence of constant background field forces us to interpret the a part of
stress energy-momentum tensor as the dark energy. Recall that cosmological
constant problem can be explained by extending Minkowski space-time to the
de Sitter space. Due to the close connection between cosmological constant
and dark energy, one can infer that dark energy can be related to the
B-gauge field.

\section*{Acknowledgements}

We would like to thank Abdurrahman Andi\c{c} for valuable discussions on the
gauge theories of gravity and the subject of extended algebras.


\begin{thebibliography}{99}
\bibitem{CM}  S. Coleman, J. Mandula, Phys. Rev. \textbf{159}, 1251-1256
(1967).

\bibitem{G}  A. Galinda, J. Math. Phys. \textbf{8}, 768 (1967).

\bibitem{BCR1}  H. Bacry, P. Combe, J.L. Richard, Nuovo Cimento \textbf{67},
267-299 (1970).

\bibitem{BCR2}  H. Bacry, P. Combe, J.L. Richard, Nuovo Cimento \textbf{70},
289-312 (1970).

\bibitem{SCH}  R. Schrader, 
Fortschritte der Physik \textbf{20}, 701 (1972).

\bibitem{BH}  J. Beckers, V. Hussin, J. Math. Phys. \textbf{24}, 1295 (1983).

\bibitem{NO1}  J.V. Negro, M.A. del Olmo, J. Math. Phys. \textbf{31}, 568
(1990).

\bibitem{NO2}  J.V. Negro, M.A. del Olmo, J. Math. Phys. \textbf{31},
2811,(1990).

\bibitem{SS1}  D.V. Soroka, V.A. Soroka, 
Phys. Lett. B \textbf{607}, 302-305 (2005).

\bibitem{SS2}  D.V. Soroka, V.A. Soroka, 
arXiv:0605251v4 [hep-th].

\bibitem{BGKL1}  S. Bonanos, J. Gomis, K. Kamimura, J. Lukierski, 
Phys. Rev. Lett. \textbf{104}, 090401 (2010).

\bibitem{GGP}  G.W. Gibbons, J. Gomis, C.N. Pope, 
Phys. Rev. D \textbf{82}, 065002 (2010).

\bibitem{BGKL2}  S. Bonanos, J. Gomis, K. Kamimura, J. Lukierski 
J. Math. Phys. \textbf{51}, 102301 (2010).

\bibitem{BGKL3}  J. Gomis, K. Kamimura, J. Lukierski 
JHEP \textbf{08}, 39 (2009).

\bibitem{FL2}  S. Fedoruk, J. Lukierski, 
JHEP \textbf{02}, 128 (2013).

\bibitem{FL3}  S. Fedoruk, J. Lukierski, 
J. Phys.: Conf. Ser. \textbf{474}, 012016 (2013).

\bibitem{AILW}  J. A. de Azcarraga, J. M. Izquierdo, J. Lukierski, M.
Woronowicz, 
Nucl. Phys. B \textbf{869}, 303-314 (2013).

\bibitem{L}  J. Lukierski, 
Proc. Steklov Inst. Math. \textbf{272}, 1-8 (2011).

\bibitem{KK}  O. Khasanov, S. Kuperstein, 
J. Phys. A: Math. Theor. \textbf{44}, 475202 (2011).

\bibitem{KL}  K. Kamimura, J. Lukierski, 
Phys. Lett. B \textbf{707}, 292-297 (2012).

\bibitem{SS3}  D.V. Soroka, V.A. Soroka, 
arXiv:1004.3194 [hep-th].

\bibitem{DG2}  R. Durka, Kowalski-Glikman, M. Szczachor, 
Mod. Phys. Lett. A \textbf{27}, 1250023 (2012).

\bibitem{HA}  S. Hoseinzadeh, A. Rezaei-Aghdam, 
arXiv:1402.0320v1 [hep-th].

\bibitem{AI}  J. A. de Azcarraga, J. M. Izquierdo, 
arXiv:1403.4128 [hep-th].

\bibitem{SS4}  D.V. Soroka, V.A. Soroka, 
Phys. Lett. B \textbf{707}, 160-162 (2012). 

\bibitem{AZKL1}  J.A. de Azcarraga, K. Kamimura, J. Lukierski, 
Phys. Rev. D \textbf{83}, 124036, (2011). 

\bibitem{DG1}  R. Durka, Kowalski-Glikman, M. Szczachor, 
Mod. Phys. Lett. A \textbf{26}, 2689-2696 (2011).

\bibitem{DG3}  R. Durka, Kowalski-Glikman, 
arXiv:1110.6812v1 [hep-th].

\bibitem{CWZ1}  S. Coleman, J. Wess, B. Zumino, Phys. Rev. \textbf{177},
2239 (1969).

\bibitem{CCWZ1}  C. Callan, S. Coleman, J. Wess, B. Zumino, Phys. Rev. 
\textbf{177}, 2247 (1969).

\bibitem{SAS1}  A. Salam, J. Strathdee, Phys. Rev. \textbf{184}, 1750 (1969).

\bibitem{SAS2}  A. Salam, J. Strathdee, Phys. Rev. \textbf{184}, 1760 (1969).

\bibitem{AIPV}  J. A. de Azcarraga, J. M. Izquierdo, M. Picon, O. Varela, 
Nucl. Phys. B \textbf{662}, 185-219 (2003).

\bibitem{LC}  L. Castellani, A. Perotto, 
Lett. Math. Phys. \textbf{38}, 321-330 (1996).

\bibitem{GGP1}  G.W. Gibbons, J. Gomis, C.N. Pope, 
Phys. Rev. D \textbf{76}, 081701 (2007).

\bibitem{BG1}  S. Bonanos, J. Gomis, 
J. Phys. A: Math. Theor. \textbf{42}, 145206 (2009).

\bibitem{BG2}  S. Bonanos, J. Gomis, 
J. Phys. A: Math. Theor. \textbf{43}, 015201 (2010). 

\bibitem{FL1}  S. Fedoruk, J. Lukierski, 
Phys. Lett. B \textbf{718}, 646-652 (2012).

\bibitem{OM}  M. Omote, 
Lettere Al Nuovo Cimento \textbf{2}, 58-60 (1971).

\bibitem{BE}  A. Bregman, 
Prog. Theor. Phys. \textbf{49}, 667-692 (1973).

\bibitem{CT}  J.M. Charap, W. Tait, 
Proc. R. Soc. Lond. A. \textbf{340}, 249-262 (1974).

\bibitem{KAS}  M. Kasuya, 
Il Nuovo Cimento B \textbf{28}, 127-37 (1975).

\bibitem{BF}  O.V. Babourova, B.N. Frolov, V.Ch. Zhukovsky, 
Phys. Rev. D \textbf{74}, 064012 (2006). 

\bibitem{Blago}  M. Blagojevic, ''Gravitation and Gauge Symmetries'',
(IoP, Bristol, 2002).

\bibitem{Weyl}  H. Weyl, ''Space, Time, Matter'', (Dover, New York, 1961).

\bibitem{BD}  C. Brans, R. H. Dicke, Phys. Rev. \textbf{124}, 925 (1962).

\bibitem{Dirac}  P.A.M. Dirac, Proc. R. Soc. Lond. A. \textbf{333}, 403-418
(1973).

\bibitem{dereli}  T. Dereli, R.W. Tucker, %
Physics Letters \textbf{110B}, 206-210 (1982). 

\bibitem{agnese}  A.G.Agnese, %
Phys. Rev. D \textbf{12}, 3804-3809 (1975). 


\end{thebibliography}

\end{document}